\magnification=1200 \vsize=25truecm \hsize=16truecm \baselineskip=0.6truecm
\parindent=1truecm \nopagenumbers \font\scap=cmcsc10 \hfuzz=0.8truecm

\font\tenmsb=msbm10
\font\sevenmsb=msbm7
\font\fivemsb=msbm5
\newfam\msbfam
\textfont\msbfam=\tenmsb
\scriptfont\msbfam=\sevenmsb
\scriptscriptfont\msbfam=\fivemsb

\def\yup{\overline y}
\def\xup{\overline x}

\def\cup{\overline c}

\def\xdo{\underline x}

\null \bigskip
\centerline{\bf A STUDY OF THE CONTINUOUS AND DISCRETE GAMBIER SYSTEMS}

\vskip 2truecm
\bigskip
\centerline{\scap A. Ramani}
\centerline{\sl CPT, Ecole Polytechnique}
\centerline{\sl CNRS, UPR 14}
\centerline{\sl 91128 Palaiseau, France}
\bigskip
\centerline{\scap B. Grammaticos}
\centerline{\sl LPN, Universit\'e Paris VII}
\centerline{\sl Tour 24-14, 5${}^{\grave eme}$\'etage}
\centerline{\sl 75251 Paris, France}
\bigskip
\centerline{\scap S. Lafortune$^{\dag}$}
\centerline{\sl LPTM et GMPIB,  Universit\'e Paris VII}
\centerline{\sl Tour 24-14, 5$^e$\'etage}
\centerline{\sl 75251 Paris, France}
\footline{\sl $^{\dag}$ Permanent address: CRM, Universit\'e de Montr\'eal,
Montr\'eal, H3C 3J7 Canada}
\bigskip\bigskip

Abstract
\smallskip \noindent  We present a systematic study of the Gambier system,
which in the continuous case is
given by two Riccati equations in cascade. We derive the condition for its
integrability and show that the
generic Gambier system contains one free function. We also derive the
Schlesinger transformations for this
system which allows in principle the systematic construction of the
integrable cases. The above procedure is
carried over to a discrete setting. We show thus how the discrete Gambier
system can be expressed as a
system of two homographic mappings in cascade. The integrable cases are
obtained through the singularity
confinement discrete integrability criterion. Finally the discrete
Schlesinger transformations are also
derived giving a handle to the construction of the integrable Gambier mapping.
\vfill\eject
\footline={\hfill\folio} \pageno=2
\noindent {\scap 1. Introduction}
\medskip Linearisable equations play a very particular role in the domain
of nonlinear systems. Their
nonlinearity is not a genuine one in the sense that elementary
transformations can reduce them to a linear
system. Calogero coined the term C-integrability (which applies to both
ordinary and partial differential
equations) in order to describe this situation and to distinguish it from
the more complicated situations of
systems integrable through IST methods (S-integrability). The simplest
example of a linearisable system is
the Riccati equation [1]:
$$x'=ax^2+bx+c\eqno(1.1)$$ which can  be transformed through the Cole-Hopf
transformation $x=-u'/(au)$ to
the linear second-order equation:
$$u''=(b+a'/a)u'-cau\eqno(1.2)$$ Higher degree, first-order linearisable
equations also exist [1]. Already
limiting ourselves to equations of binomial type  we find
$$(x')^2=(x-a)^2(x-k_1)(x-k_2)\eqno(1.3)$$ where $a$ is a function of the
independent variable and $k_1,k_2$
are constants. Putting $u^2=(x-k_1)/(x-k_2)$ equation (1.3) reduces to the
Riccati:
$$u'=\pm{1\over 2}(k_1-a-(k_2-a)u^2)\eqno(1.4)$$ When we consider
second-order equations, the situation
becomes immediately richer, in the sense that there exist several types of
linearisable equation [1]. The
first (and simplest) one is the equation obtained if one computes the
derivative of both sides of the Riccati
(1.1). It is usually given in canonical form and reads:
$$x''=-2xx'+bx'+b'x\eqno(1.5)$$  In the same spirit, one can compute the
derivative of (1.1) after having
divided both sides of the equation by $x$. We find thus the equation (again
given in canonical form):
$$x''={x'^2\over x}+\left(ax-{c\over x}\right) x'+a'x^2+c'\eqno(1.6)$$
Another linearisable equation does
exist which can be linearized by a Cole-Hopf transformation just like the
Riccati. Its form is:
$$x''=-3xx'-x^3+q(x'+x^2)\eqno(1.7)$$  Indeed, putting $x=u'/u$ we can
reduce it to the linear third order
equation:
$$u'''=qu''\eqno(1.8)$$

But the most interesting linearisable equation discovered at second-order
is the one obtained by Gambier in
his classification of second-order equations having the Painlev\'e property
[2]. The Gambier equation is
usually given as a second-order equation for a single variable:
$$\matrix{
\displaystyle{x''={n-1\over n}{x'^2\over x}+a{n+2\over n}xx' +bx'
-{n-2\over n}{x'\over x}\sigma-{a^2 \over
n}x^3 + (a'-ab)x^2} \cr
\displaystyle{+ \Big(cn-{2a\sigma\over n}\Big)x-b\sigma-{\sigma^2\over nx}}}
\eqno(1.9)$$ where $n$ is integer, $a,b,c$ are functions of the independent
variable and $\sigma$ is equal
to 0 or 1. Its structure becomes clearer when one writes it as a system. We
have in this case a system of two
Riccati's:
$$y'=-y^2+by+c\eqno(1.10a)$$
$$x'=ax^2+nxy+\sigma\eqno(1.10b)$$ Although the Gambier equation is always
linearisable this does not mean
it is always integrable.  Indeed, in a system such as (1.10) of two
equations in cascade we can always solve
the first equation for $y$, obtain $y(t)$ and inject it into the second
equation. The latter can always be
written as a linear, second-order differential equation for $x$. So, in
principle, the problem can always be
solved formally. The difficulty comes when one wishes to actually compute
$x$, in terms of contour integrals, while $y$ has an analytic structure
that interferes badly with that of
$x$. This is where the Painlev\'e property comes into play. If we require
that the system possess the
Painlev\'e property the integration can be performed and we can indeed
obtain the solution for $x(t)$ over
the complex $t$-plane. Thus, the integrability of the Gambier equation will
be closely related to its
singularity structure.

The fact that the integrability of the Gambier equation is related to the
Painlev\'e property allows us to
obtain another interesting result. Just as in the case of the Painlev\'e
equations, it is possible to
introduce transformations relating the solution of an equation with
parameter $n$ to one with parameter
$n+1$ or $n+2$ (two different transformations do exist). Thus the Gambier
system possesses Schlesinger
transformations [3].

Another feature of the Gambier equation is that it can be integrably
discretized [4,8]. Our approach follows
closely the spirit of Gambier based on coupled Riccati equations. In
perfect analogy to the continuous case
it is possible to introduce a system of two homographic mappings in cascade
which represent the
discretization of the Gambier system. The general form of this system is
the following:
$$y_{n+1}={ay_n+b\over cy_n+d}\eqno(1.11a)$$
$$x_{n+1}={(\alpha y_n+\beta)x_n+\gamma y_n+\delta\over (\epsilon
y_n+\zeta)x_n+\eta
y_n+\theta}\eqno(1.11b)$$ The form (1.11) can be simplified through
homographic transformations and the
integrable cases can be obtained through the application of the singularity
confinement discrete
integrability criterion. As in the continuous case, although (1.11) is
always linearisable it is not
automatically integrable. The difficulty arises when one tries to compute
$x_n$ in terms of matrix products
the elements of which contain $y_n$. Some of these matrices are singular in
such a way that degrees of
freedom are irretrievably lost when $y_n$ has the wrong properties.
Singularity confinement precisely means
that these degrees of freedom  are in fact recovered at some later stage.

Quite expectedly, the Gambier mapping has Schlesinger transformations.
Their derivation is based on a study
of the singularities of the mapping (1.11) and their confinement. Thus the
parallel between the discrete and
continuous cases is perfect.

In what follows, we shall present the singularity analysis of the Gambier
equation and derive its integrable
cases. Next, we study the discrete case and use the singularity structure
in order to derive the Gambier
mapping. Finally, we present the Schlesinger transformations of both the
continuous and discrete Gambier
systems which allow us, starting from some elementary case ($n$=0 or
$n$=1), to construct recursively the
Gambier systems for higher $n$'s.
\bigskip
\noindent {\scap 2. The continuous Gambier equation}
\medskip The Gambier equation is given as a system of two Riccati equations
in cascade. This means that we
start with a first Riccati for some variable $y$
$$ y'=-y^2+by+c \eqno(2.1)$$  and then couple its solution to a second
Riccati by making the coefficients of
the latter depend explicitly on $y$:
$$ x'=ax^2+nxy+\sigma. \eqno(2.2)$$ The precise form of the coupling
introduced in (2.2) is due to
integrability requirements. In fact, the application of singularity
analysis shows that the Gambier system
cannot be integrable unless the coefficient of the $xy$ term in (2.2) is an
integer $n$. This is not the
only integrability requirement. Depending on the value of $n$ one can find
constraints on the $a$, $b$, $c$,
$\sigma$ (where the latter is traditionnally taken to be constant $1$ or
$0$) which are necessary for
integrability.

The common lore [1] is that out of the functions $a$, $b$, $c$ two are
free. This turns out not to be the
case. The reason for this is that the system (2.1-2) is not exactly
canonical i.e. we have not used all
possible transformations in order to reduce its form. We introduce a change
of independent variable from $t$
to $T$ through
$dt=gdT$ where
$g$ is given by ${1\over g}{dg \over dt}=b{n\over 2-n}$, a gauge through
$x=gX$ and also
$Y=gy-{1\over n}{dg\over dt}$. The net result is that system (2.1-2)
reduces to one where $b=0$ while
$\sigma$ remains equal to $0$ or $1$. It is clear from the equations above
that $n$ must be different from
$2$. On the other hand when $n=2$ the integrability conditions, if
$\sigma=1$, is precisely
$b=0$. So we can always take $b=0$. (As a matter of fact in the case
$\sigma=0$ an additional gauge freedom
allows us to take both $b$ and $c$ to zero for all $n$, even for $n=2$).
Thus the Gambier system can be
written in full generality
$$ y'=-y^2+c \eqno(2.3a)$$
$$ x'=ax^2+nxy+\sigma. \eqno(2.3b)$$

One further remark is in order here. The system (2.3) retains its form
under the transformation
$x\rightarrow 1/x$. In this case $n \rightarrow -n$ and $\sigma$ and $-a$
are exchanged. Thus in some cases
it will be interesting to consider a Gambier system where $\sigma$ is not
constant but rather a function of
$t$. Still, it is possible to show that we can always reduce this case to
one where
$\sigma=1$, while preserving the form of (2.3a) i.e. $b=0$. To this end we
introduce the change of variables
$dt=hdT$, $x=gX$ and $Y=hy-{1\over 2}{dh\over dt}$ with $h=\sigma^{2/(n-2)}$,
$g=\sigma^{n/(n-2)}$. With these transformations system (2.3) reduces to
one with $\sigma=1$ and $b=0$. (In
the special case $n=2$, with $b=0$, integrability implies
$\sigma=$constant, whereupon its value can always
be reduced to $1$).

In order to study the movable singularities  of the coupled Riccati system
we start from the observation
that from (2.3a) the dominant behaviour of  $y$  can only be $y\approx
{1/(t-t_0)}$. The next terms in the
expansion of $y$ can be easily obtained, and involve the function  $c$ and
its derivatives.  In order to
study the structure of the singularities of (2.3b), we first remark that
since the latter is a Riccati, its
movable singularities are poles. However, (2.3b) also has singularities
that are due to the singular
behaviour of the coefficients of the r.h.s of (2.3b), namely $y$.  Now, the
locations of the singularities
of the coefficients are `fixed' as far as (2.3b) is concerned. However,
from the point of view of the full
system(2.3), these singularities are {\sl movable} and thus should be
studied. The `fixed' character
reflects itself in the fact -1 is {\sl not} a resonance. (The terms
`resonance' is used here following the
ARS terminology [6] and means the order, in the expansion, where a free
coefficient enters. A resonance -1
is related to the arbitrariness of the location of the singularity, and is
thus absent when the location of
the singularity is determined from the `outside' rather than by the initial
conditions). Because of the pole
in $y$, $x$ has a singular expansion with a resonance different from -1
which may introduce a compatibility
condition to be satisfied.

We consider below the case of the full Riccati (2.3b) with $a\neq 0$,
$\sigma=1$. (The analysis of the case
$a\sigma=0$ was given in [4]). As we explained above, only the singularity
due to
$y$ can lead to trouble. Rewriting (2.3b) as
${x'/ x}=ax+ny+{\sigma/x}$ for $y=1/(t-t_0)+\dots$ we remark that unless
$n=\pm 1$  a behaviour of the form
$x\sim (t-t_0)^n$ is impossible when $a\sigma\neq 0$.  For $n=1$, a
logarithmic leading behaviour will be
present for $\sigma\neq 0$. (Note that the condition $\sigma=0$ is
sufficient for the absence of a critical
singularity for $n=1$ irrespective of the value of $a$. Similarly, in a
dual way, for $n=-1$ the necessary
and sufficient condition for the absence of a critical singularity is
$a=0$,  whether $\sigma$ is 1 or 0).

Next we assume $n\neq \pm 1$, in which case it suffices to study the
singularities
$x\approx
\lambda (t-t_0)$, $(\lambda=\sigma /(1-n))$   and $x\approx
\mu/(t-t_0)$, $(\mu=-(n+1)/a)$. The first singular behaviour ($x\approx
\lambda (t-t_0)$) has a resonance at $n-1$,  which is negative for $n<1$
and thus does not introduce any
further condition.  For $n>1$, the resonance condition can be studied at
least for the first few values of
$n$. (In fact, $\sigma=0$ suffices for the resonance condition to be
satisfied even for $a\neq 0$). In the
particular case $n=2$, we have already mentioned that the integrability
condition for $\sigma=1$ is precisely
$b=0$. For higher $n$ (and $\sigma=1$) we find the further possibilities:
$$n=3\quad\quad 2c-a=0\eqno(2.5)$$
$$n=4\quad\quad  3c'-a'=0$$

The second singular behaviour, $x\approx \mu/(t-t_0)$, has a resonance at
$-1-n$.  Thus for $n>0$ this
resonance is negative and does not introduce any further condition, while
for  $n<0$ a compatibility
condition must be satisfied.  (We find that for every case $n<0$, $a=0$ is
a sufficient condition for the
absence of critical singularity. This is not in the least astonishing given
the duality of $a$ and
$\sigma$). On the other hand if we demand
$a\neq 0$ then a different resonance condition is obtained, at each value
of $n$.  For $n=-1$, whenever
$a\neq 0$, a logarithmic singularity of the form $(t-t_0)^{-1}\log (t-t_0)
$ appears irrespective of the
value of
$\sigma$. For $n<-1$, we find:
$$n=\!-2\quad  a'=0 $$
$$n=\!-3\quad 2ac-a^2\sigma-2a''=0\quad\eqno(2.6)$$
$$n=\!-4\quad 2ac'+a'(4c-2a\sigma)-a'''=0$$  The integrability condition
for higher values of $n$ can be
obtained through the use of computer algebra.
\bigskip
\noindent {\scap 3. The discrete Gambier equation}
\medskip The discretisation of the Gambier equation is based on the idea of
two Riccati equations in
cascade. The discrete form of the first is simply:
$$\yup={ay+b\over cy+d}\eqno(3.1)$$ where $y\equiv y_n$ and $\yup\equiv
y_{n+1}$. The second equation which
contains the coupling can be discretised in several, not necessarily
equivalent, ways. In [5] we have
considered the
 generic coupling of the form:
$$\alpha x\xup y+\beta x\xup+\gamma\xup y+\delta\xup+\epsilon xy+\zeta
x+\eta y+\theta=0\eqno(3.2)$$
Implementing a homographic transformation on $x$ and $y$ we can generically
bring (3.2) under the form:
$$x\xup+\gamma\xup y-\epsilon xy-\theta=0\eqno(3.3)$$. A choice of
different transformations can bring
(3.2) also to the form
$$\xup-x=-fx\xup+(gx+h\xup)y+k\eqno(3.4)$$ Note that (3.2) contains an
`additive' type coupling
$x\xup+\delta\xup+\zeta x+\eta y+\theta=0$ for special values of its
parameters, but the generic form (3.3)
is that of a `multiplicative' coupling where $\gamma,\epsilon$ do not
vanish. Solving (3.3) for
$\xup$ we obtained the second equation of the discrete Gambier system in
the form:
$$\xup={\epsilon xy+\theta\over x+\gamma y}.\eqno(3.5)$$ Clearly, a scaling
freedom remains in equation
(3.5). We can use it in order to bring it to the final form:
$$\xup={xy/d+c^2 \over x+dy} \eqno(3.6)$$ Eliminating $y$ and $\yup$ from
(3.1), (3.6) and its upshift, we
can obtain a 3-point mapping for $x$ alone but the analysis is clearer if
we deal with both $y$ and
$x$.

In what follows we shall present a different derivation based on the study
of singularities of the system.
As we explained in [3] we can use the  homographic freedom in order to
bring the mapping for $y$ to the form:
$$\yup={y+c \over y+1} \eqno(3.7)$$ instead of (3.1) where $c$ is a
function of $n$. Next, we turn  to the
equation for $x$. This equation is homographic in $x$. However we require
that when
$y$ takes the value $0$, the resulting value of $x$ be $\infty$. Thus the
denominator must be proportional
to $y$, and since we can freely translate $x$, we can reduce its form to
just $xy$. The remaining overall
gauge factor is chosen so as to put the coefficient of $xy$ of the
numerator to unity resulting to the
following mapping:
$$
\xup={x(y-r)+q(y-s) \over xy}. \eqno(3.8)
$$ The system (3.7-8) is a discrete form of the Gambier system. In order to
study the confinement of the
singularity induced by $y=0$ we introduce the auxiliary quantity $\psi_N$
which is the $N$'th iterate of
$y=0$ in equation (3.7), $N$ times downshifted. Thus $\psi_0=0$,
$\psi_1=\underline{c}$,
$\psi_2={\underline{\underline{c}}+\underline{c} \over
\underline{\underline{c}} +1}$, etc... The
confinement requirement is that after $N$ steps $x$ becomes $0$ in such a
way as to lead to $0/0$ at the
next step. Thus the mapping (3.8) has in fact the form:
$$\xup={x(y-r)+q(y-\psi_N) \over xy}. \eqno(3.9)$$  Thus when at some step
$N$ we have $y=\psi_N$ and $x=0$,
on the view of (3.9)  $\xup$ will then  be indeterminate of the form
$0/0$. However it turns out that in fact this value is well-determined and
finite. Let us take a closer look
at the conditions for confinement. The generic patterns for $x$ and $y$ are:
$$
\matrix{
\displaystyle{ y: \{} & & \displaystyle{0}&& \displaystyle{\overline{\psi_1}}&&
\displaystyle{\overline{\overline{\psi_2}}}&& \displaystyle{\dots}
&&\displaystyle{\overline{\dot{\dot{\overline{\overline{{\psi}}}}}}_N}&&
\} \cr
\displaystyle{x: \{}& {\rm free}&& \infty &&
\displaystyle{{\overline{\psi_1}-\overline{r} \over
\overline{\psi_1}}}&&\dots && 0 &&{\rm free}& \ \}\hfill. }
$$ At $N=1$ it is clearly impossible to confine with a form (3.9) since we
do not have enough steps. In this
case the only integrable form of the $x$-equation is a linear one.
 This case must be studed
separately. For an arbitrary
$N$, the general form of the linear $x$-mapping can be obtained using
confinement arguments in a way similar
to what we did for the generic, nonlinear, case. We obtain
$$
\xup={x(y-\psi_N)+g \over y} \eqno(3.10)
$$ where $g$ is free.

The first genuinely confining case of the
form (3.9) is $N=2$. From the requirement $\overline{\xup}=0$ we have
$r=\psi_1$ and $q$ free: this is
indeed the only integrability condition. For higher $N$'s we can similarly
obtain the confinement condition
which takes the form of an equation for $r$ in terms of $q$.

At this point it is natural to ask whether the mapping (3.7)-(3.9) does
indeed correspond to the Gambier
equation (2.3). In order to do this we construct its continuous limit. We
first introduce:
$$
\matrix{
\displaystyle{c=\epsilon^2D} \cr \cr
\displaystyle{y={\epsilon D \over Y+H}} }
\eqno(3.11)
$$  with $H\approx D'/(2D)$ and obtain the continuous limit of (3.7) for
$\epsilon \to 0$. We find as
expected
$$ Y'=-Y^2+C \eqno(3.12)
$$ (i.e. eq. (2.3a)) where
$C=D-{D''\over 2D}+{3\over 4}{D'^2 \over D^2}$. Using (3.7) and (3.11) we
can also compute $\psi_N$ and we
find at lowest order:
$$
\psi_N=\epsilon^2\Psi_N
\qquad{\rm with }\qquad
\Psi_N\approx N(D-\epsilon{N+1 \over 2}D')+\epsilon^2\Phi_N \eqno(3.13)
$$ where $\Phi_N$ is an explicit function of $D$ depending on $N$.

Next we turn to the equation for $x$ and introduce:
$$
\matrix{
\displaystyle{r=\epsilon^2 R} \cr \cr
\displaystyle{x={1\over 2}+{\epsilon \over 2X}-\epsilon {R D' \over 4D^2}}
\cr \cr
\displaystyle{q\approx-{1\over 4}+\epsilon^2Q} }
\eqno(3.14)
$$ and for the continuous limit of the form (2.3b) to exist in canonical
form (i.e. $b$=0, $\sigma$=1) we
find that we must have
$$ R\approx{ND\over 2} -\epsilon (N+2){ND' \over 8}.
\eqno(3.15)$$  This leads to the equation for $x$:
$$ X'=AX^2+NXY+1 \eqno(3.16)
$$ with $A={N\over 4}(N/4+1){D'^2 \over D^2}-{ND'' \over 4D}-4Q$. Moreover
the confinement constraint
implies a differential relation between $D$ and $Q$ which depends on $N$.
We can verify explicitly in the
first few cases that this is indeed the integrability constraint obtained
in the continuous case. In the
case $N=2$, the condition for confinement is
$$
\overline{r} =c. \eqno(3.17)
$$ Using the approximate expansion in $\epsilon$ (3.15) which gives $r$ up
to the third order in $\epsilon$
and putting
$N=2$, we get that equation (3.17) is automaticaly satisfied for the two
first nonzero orders in $\epsilon$
($\epsilon^2$ and
$\epsilon^3$). This reflects the fact that for the continuous Gambier
equation, in the case $N=2$, the only
integrability condition is
$b=0$. Thus no condition has to be imposed on $D$ and
$Q$.

The confinement condition for
$N=3$ is obtained by imposing $\overline{\overline{\xup}}=0$ when $y=0$.
This condition reads:
$$ (\cup
+c-(\cup+1)\overline{\overline{\psi_3}})=(\overline{r}-c)
(\cup+c-\overline{\overline{r}}(\cup+1)).
\eqno(3.18)
$$ Putting $N=3$, we find that equation (3.18) is identically satisfied for
the first two nonzero orders in
$\epsilon$ ($\epsilon^4$ and
$\epsilon^5$). To get the integrability condition, we must calculate (3.18)
at order $\epsilon^6$. To do
this, we need the value of $\Phi_3$ in (3.13). We easily find that
$\Phi_3=7D''-8D^2$. Satisfying (3.18) in $\epsilon^5$ gives a relation
giving explicitly $Q$ in terms of
$D$. Implementing this relation, we find that $A=2C$ which is the
continuous integrability condition for
$N=3$.
\bigskip
\noindent {\scap 4. Schlesinger transformations for the continuous Gambier
equation}
\medskip

The theory of auto-B\"acklund transformations of Painlev\'e equations is
well established. As was shown in
[7] the general form of auto-B\"acklund transformations for most Painlev\'e
equations is of the form:
$$
\tilde{x}={\alpha x'+\beta x^2+\gamma x+\delta \over \epsilon x'+\zeta
x^2+\eta x+\theta }.
\eqno(4.1)
$$ In the case of the Gambier equation considered as a coupled system of
two Riccati's it is more convenient
to look for an auto-B\"acklund of the form:
$$
\tilde{x}={\alpha xy+\beta x+\gamma y+\delta \over  (\zeta y+\eta)(\theta x
+\kappa)}.
\eqno(4.2)
$$ with a factorized denominator, with hindsight from the discrete case. We
require that the equation
satisfied by $\tilde x$ do not comprise terms nonlinear in $y$. We examine
first the case $\zeta\neq 0$ and
reach easily the conclusion that there exists no solution. So we take
$\zeta =0$, $\eta=1$ which implies
that $\alpha$ and $\gamma$ do not both vanish (otherwise (4.2) would have
been independent of $y$). We find
in this case $\alpha=0$ and thus the general form of the auto-B\"acklund
can be written as:
$$
\tilde{x}={\beta x+\gamma y+\delta \over \theta x+\kappa}. \eqno(4.3)
$$ From (4.3) we can obtain the two possible forms of the Gambier system
auto-B\"acklund:
$$
\tilde{x}=\beta x+\gamma y+\delta \eqno(4.4)
$$
$$
\tilde{x}={\beta x+\gamma y+\delta \over x+\kappa}. \eqno(4.5)
$$
 As we shall see in what follows both forms lead to Schlesinger
transformations.

Let us first work with form (4.4). Our approach is straightforward. We
assume (4.4) and require that
$\tilde x$ satisfy an equation of the form (2.3b) while $y$ is always the
same solution of (2.3a). The
calculation is easily performed leading to:
$$
\tilde{x}=\gamma y+{a \gamma \over n+1}x+{\gamma' \over n}, \eqno(4.6)
$$ where $\gamma$ satisfies:
$$ {\gamma' \over \gamma}={n \over n+2}{a' \over a}. \eqno(4.7)
$$ Here we have assumed $a\neq 0$; otherwise $\tilde x$ does not depend on
$x$ and (4.6) does not define a
Schlesinger. The parameters of the equation satisfied by
$\tilde x$ are given (in obvious notations) by:
$$
\tilde{n}+n+2=0 \eqno(4.8a)
$$
$$
\tilde{a}={n+1 \over \gamma} \eqno(4.8b)
$$ and
$$
\tilde{\sigma}=\gamma\Big(c+{a\sigma \over n+1}+{1\over n+2}{a''\over
a}-{n+3\over (n+2)^2}{a'^2\over
a^2}\Big).
\eqno(4.8c)
$$ Thus (4.6) is indeed a Schlesinger transformation since it takes us from
a Gambier system with parameter
$n$ to one with parameter $\tilde n=-n-2$. It suffices now to invert
$\tilde x$ in order to obtain an
equation with parameter $N=n+2$. Expressions (4.6) and (4.8) can be written
in a more symmetric way:
$$
\tilde{a}\tilde{x}-a x=(n+1)(y-{a' \over \tilde{n}a}) \eqno(4.9)
$$ and
$$
\matrix{
\displaystyle{\tilde{n}+1=-(n+1)}\cr
\displaystyle{\tilde{n}{\tilde{a}' \over \tilde{a}}=n{a' \over a}}\cr
\displaystyle{\tilde{a}\tilde{\sigma}-a\sigma=(n+1)\Big(c-{1\over \tilde{n}}
\big({a'\over a}\big)'+{1\over\tilde{n}^2}{a'^2\over a^2}\Big). } }
\eqno(4.10)
$$

The inverse transformation can be easily obtained if we introduce $\tilde
\gamma$ such that
$a\tilde{\gamma}=-(n+1)=-\tilde{a}\gamma$. We thus find
$$ x=y \tilde{\gamma}+{\tilde{a}\tilde{\gamma}\over
\tilde{n}+1}\tilde{x}+{\tilde{\gamma}' \over \tilde{n}}
\eqno(4.11)
$$ and the relations (4.10) are still valid.

Iterating the Schlesinger transformations one can construct the integrable
Gambier systems for higher
$n$'s and obtain by construction the functions which appear in them.
However it may happen that when we
implement the Schlesinger we find
$\tilde{\sigma}=0$. If we invert $x$ we get a system with
$N=-\tilde{n}=n+2$ but $A=0$ for which one cannot
iterate the Schlesinger.

Let us give example of the application of this Schlesinger transformation.
Let us start from $n=0$, in which
case we find $\tilde{n}=-2$ and, after inversion, $N=2$. For $n=0$ we start
from $a=-1$ and
$\sigma=0$ or $1$ (always possible through the appropriate changes of
variable). This leads to
$\tilde{a}=-1$, $\tilde{\sigma}=-c+\sigma$ and the Schlesinger reads:
$\tilde{x}=-y+x$. Next we invert
$\tilde{x}$ and have $X=1/(x-y)$. We find thus that the Schlesinger takes
us from
$$
\matrix{ y'=-y^2+c \cr x'=-x^2+\sigma }
\eqno(4.12)
$$ to the system
$$
\matrix{ y'=-y^2+c \cr X'=AX^2+2Xy+\Sigma }
\eqno(4.13)
$$ with $A=c-\sigma$, $\Sigma=1$. In the particular case $n=2$, a change of
variables exists which allows us
to put $A=-1$ (unless $A=0$), without introducing $b$ in the equation for
$y$, while keeping $\Sigma=1$ and
changing only the value of
$c$. Thus the generic case of the Gambier equation for $n=2$ can be written
with $A=-1$. Eliminating
$y$ between the two equations we find:
$$ x''={x'^2 \over 2x}-2xx'-{x^3 \over 2}-{1 \over 2x}+(2c+1)x. \eqno(4.14)
$$  This is the generic form of the $n=2$, Gambier equation [1] and it
contains just one free function.

We turn now to the second Schlesinger transformation corresponding to the
form (4.5). As we shall show, a
Schlesinger transformation of this form does indeed exist and corresponds
to changes in $n$ with $\Delta
n=1$. Let us start from the basic equations (2.3). Next we ask that $\tilde
x$ defined by (4.5) indeed
satisfy a system like (2.3). We find thus that and $\kappa=-x_0$ and
$\gamma$ must be given by:
$$ {\gamma' \over \gamma}=y_0+{2ax_0 \over n+1} \eqno(4.15)
$$ where
$y_0$ is a solution of the Riccati (2.3a) and a solution $x_0$ of (2.3b),
obtained with $y$ replaced by
$y_0$. We introduce the quantities $\tilde{x}_0={a\gamma
\over n+1}$,
$\tilde{a}=-{n x_0\over \gamma}$. In this case (4.15) becomes:
$$ {\gamma' \over \gamma}=y_0+{2\tilde{a}\tilde{x}_0 \over
\tilde{n}+1}=y_0+{2x_0\tilde{x}_0 \over \gamma},
\eqno(4.16)
$$ where
$$
\tilde{n}+n+1=0. \eqno(4.17)
$$  We have thus, starting from a generic solution $x$, $y$ of (2.3) for
some $n$, the Schlesinger:
$$
\tilde{x}=\tilde{x}_0+{\gamma(y-y_0) \over x-x_0} \eqno(4.18)
$$ where $\tilde{x}$  is indeed a solution of (2.3) for $\tilde{n}=-n-1$
for the {\sl same} $y$
$$\tilde{x}'=\tilde{a}\tilde{x}^2+\tilde{n}\tilde{x}y+\tilde{\sigma}
\eqno(4.19)
$$ 
where $\tilde{a}$  has
been defined as $-nx_0/\gamma $ and
$$\tilde{\sigma}={\gamma \over n+1}\Big(a'+a^2x_0{n+2\over
n+1}+ay_0(n+2)\Big). \eqno(4.20)$$  Note that
$\tilde{x}_0$ is a solution of the same equation with $y$ replaced by
$y_0$. As in the previous case if we
invert $\tilde x$ we obtain an equation corresponding to $N=n+1$.

It is worth pointing out here that the Schlesinger transformation
corresponding to $\Delta n=2$ was known to
Gambier himself. As a matter of fact when faced with the problem of
determining the functions appearing in
his equation so as to satisfy the integrability requirement, Gambier
proposed a recursive method which is
essentially the Schlesinger $\Delta n=2$. On the other hand the Schlesinger
$\Delta n=1$ is quite new.
\bigskip
\noindent {\scap 5. Schlesinger transformations for the discrete Gambier
equation}
\medskip
 Once the singularity pattern of the Gambier mapping is established we can
use it in order to construct the
Schlesinger transformation. Let us first look for a transformation that
corresponds to
$\Delta N=2$. The idea is that given the $N$-steps singularity pattern of
the equation for $x$ we introduce a
variable $w$ with $N+2$ singularity steps where we enter the singularity
one step before $x$ and exit it one
step later. The general form of the Schlesinger transformation, which
defines $w$, is:
$$ w=X{y-\psi_{N+1} \over y}, \eqno(5.1)
$$ where $X$ is homographic in $x$. The presence of the $y$ and
$y-\psi_{N+1}$ terms is clear: they ensure
that $w$ becomes infinite one step before $x$, and vanishes one step after
$x$. Next we turn to the
determination of $X$. Since $X$ is homographic in $x$ we can rewrite (5.1) as:
$$ w={\alpha x+\beta \over y}{y-\psi_{N+1} \over \gamma x+\delta}. \eqno(5.2)
$$
 Our requirement is that $w$  becomes infinite when $y=0$ for every value
of $x$. This statement must be
qualified. The numerator $\alpha x+\beta$ {\it will} vanish for some $x$
(namely
$x=-\beta/\alpha$) so this value of $x$ must be the only one which should
{\it not} occur in the confined
singularity. Indeed there is a unique value of $x$ where instead of being
confined, the singularity extends
to infinity in {\it both} directions of the independent variable $n$, while
the only nonsingular values of
the dependent variable occur in a finite range. The value of $x$ such that
$\xup$ is finite and free even
though $y$ is zero is such that the numerator $-xr-q\psi_{N}$ of  $\xup$
vanishes.   For this value of $x$,
the values of the dependent variable are fixed for $n\le 0$  and $n\ge N+1$
and the value can be considered
as `forbidden'. Thus $\alpha x +\beta=xr+q\psi_{N}$ up to a multiplicative
constant. Similarly when
$y=\psi_{N+1}$, $w$ must vanish. Thus
$\gamma x+\delta$ must not be zero except for the unique value of $x$ that
does not occur in the confined
singularity. Note that $y=\psi_{N+1}$ means $\underline y=\underline
{\psi}_N$ and the only value of $x$
that comes from a nonzero $\xdo$ in that case is
$x={(\underline{\psi}_N-\underline r) /\underline{\psi}_N}$. In that case
the values of the dependent
variable are fixed for $n\ge 0$ and $n\le -N-1$. This value of $x$ being
`forbidden',
$\gamma x+\delta$ must be proportional to
$\underline{\psi}_N x-(\underline{\psi}_N-\underline{r})$. We now have the
first form of the Schlesinger:
$$ w={xr+q\psi_N \over y}{y-\psi_{N+1} \over \underline{\psi}_N x
+\underline{r}-\underline{\psi}_N}
\eqno(5.3)$$
where the proportionality constant has been taken equal to $1$ (but any
other value would have been
equally acceptable). Here $w$ effectively depends on $x$ unless
$r(\underline{r}-\underline{\psi}_N)=q\underline{\psi}_N\psi_N$. But in
this case the mapping (3.8) is in
fact linear in the variable
$\xi=(x-1+\overline{r}/\overline{\psi}_N)^{-1}$. This case is the analog of
the
case $a=0$ in the continuous case where the Schlesinger does not exist. Let
us give an application of the
Schlesinger transformation by obtaining the
$N=2$ equation starting from $N=0$. We have always the equation for $y$
which reads:
$$
\yup={ y+c \over y+1} \eqno(5.4)
$$ and $\psi_0=0$, $\psi_1=\underline{c}$. For $N=0$ the equation for $x$
reads:
$$
\xup={x(y-r)+qy \over xy}={x+q \over x} \eqno(5.5)
$$  since for integrability $r=0$ and indeed $N=0$ means that the $x$
equation does {\sl not} depend on $y$.
We introduce the Schlesinger:
$$ w=x{y-\underline{c} \over y} \eqno(5.6)
$$ Using (5.5) and (5.6) to eliminate $x$ we obtain the equation for $w$:
$$
\overline{w}=(1-c){yw+q(y-\underline{c}) \over (y+c)w}. \eqno(5.7)$$
This equation is of the form (3.8) but
not quite canonical. We can transform it to canonical form simply by
introducing $\yup$ instead of $y$
because indeed $w$ is infinite one step before $x$, so $w=\infty$ means
$\xup=\infty$ i.e. $\yup=0$. We obtain thus:
$$
\overline{w}={w(\yup-c)+q(1+\underline{c})(\yup-\overline{\psi}_2) \over
\yup w} \eqno(5.8)
$$
with $\overline{\psi}_2=(c+\underline{c})/(1+\underline{c})$ which coupled
to (5.4) is indeed a $N=2$
Gambier mapping.

 As we pointed out in section 3, necessitates a special treatment. The
Schlesinger transformation is again
given by:
$$ w=X{y-\psi_{N+1} \over y} \eqno(5.9)
$$ and arguments similar to those of the nonlinear case allow us to
determine the form of the homographic
object $X$ leading to:
$$ w={x\psi_N-g \over y}{y-\psi_{N+1} \over
x\underline{\psi}_N-\underline{g}}. \eqno(5.10)
$$
Thus one can also perform a Schlesinger in the linear case. This is not in
disagrement with the
continuous case. It is, in fact, the analog of the case where $\sigma=0$
but $a\neq 0$ (which is linear in
$1/x$) for which the Schlesinger {\sl can} be performed. The analog of the
case $\sigma=0$ and $a=0$ is the
situation when
$g=k\psi_N$ with constant
$k$ in which case  the mapping rewrites $\overline{\xi}=\xi(y-\psi_N)/y$
with $\xi=x-k$. Then $w$ does not
depend on $\xi$ (or $x$) and (4.20) does not define a Schlesinger in
analogy to the case
$r(\underline{r}-\underline{\psi}_N)=q\underline{\psi}_N\psi_N$ in the
nonlinear case.

Finally, we examine the possibility of the existence of a $\Delta N=1$
Schlesinger. In this case, the
structure of the transformation will be obtained by asking that the $N+1$
case enter the singularity one
step before the $N$ case but exit at the same point. The general structure
is thus:
$$ w={rx+q\psi_N \over y}{y-\eta \over x-\xi} \eqno(5.11)
$$ where $\eta$ and $\xi$ must be determined. We do this by requiring that
the equation for $w$ contain no
coefficients nonlinear in $y$. As a result we find that $\eta$ must satisfy
the equation (3.7) for $y$:
$$
\overline{\eta}={\eta+c \over \eta+1} \eqno(5.12)
$$ and $\xi$ the equation (3.9) for $x$ with $\eta$ instead of $y$:
$$
\overline{\xi}={\xi(\eta-r)+q(\eta-\psi_N) \over \xi \eta}. \eqno(5.13)
$$ We remark here the perfect parallel to the continuous case (and as we
pointed out the discrete case led
the investigation back to the continuous one). Let us point out here that
the $w$ obtained through (4.26)
does not lead to $w=0$ at the exit of the singularity (i.e. when $x=0$,
$y=\psi_N$) and a translation is
needed. One has in principle to define a new variable
$$
\omega = w-w(x=0,y=\psi_N)=w+{q \over \xi}(\psi_N-\eta).
$$

Finally we derive the $\Delta N=1$ Schlesinger for the case of a linear
mapping (5.9). We start from:
$$ w={x\psi_N-g \over y}{y-\eta \over x-\xi} \eqno(5.14)
$$ and again require for $w$ an equation with coefficients linear in $y$.
We find that $\eta$ must again be
a solution of the equation for $y$ i.e. it must satisfy (5.12) and moreover
$\xi$ is a solution of (5.13)
with $y=\eta$:
$$
\overline{\xi}={\xi(\eta-\psi_N)+g \over \eta}. \eqno(5.15)
$$ Thus the list of the Schlesinger transformations of the Gambier mapping
is complete.
\bigskip
\noindent {\scap 6. Conclusion}
\medskip

In this work we have reviewed the Gambier system in both its continuous and
discrete forms. We have shown that the singularity analysis can be used in
order to obtain  the integrable
subcases of this equation and moreover we have derived their
Schlesinger transformations.
Several open questions appear at this point. The Gambier equation is the
generic linearisable second-order
equation of first degree. If one relaxes this last constraint, one can
already obtain further linearisable
equations. Cosgrove and Scoufis [8] have presented two such examples:
$$(x'')^2=(ax'+a'x+c')^2x' \eqno(6.1)$$
$$
(x'')^2=A(x)(c_1(tx'-x)+c_2x'(tx'-x)+c_3(x')^2+c_4(tx'-x)+c_5x'+c_6).
\eqno(6.2)$$
It would be interesting to study closely the properties of these equations
and in particular their
discretization.

Moving to higher orders one can wonder how the Gambier approach can be
extended. In [9] we have presented a
first exploration of this problem, in both continuous and discrete
settings, for third-order systems.

Finally, the general problem of  $n$-th-order linearizable systems is far
from being solved. Only a special
class of such systems is known, based on projective contructions. Recently,
we have obtained the
discretization of these projective Riccati systems [10]. Clearly, this is
not the last word as far as
linearisability is concerned.

 \bigskip
\noindent {\scap Acknowledgements}.
\smallskip
\noindent
 S. Lafortune acknowledges two scholarships: one from NSERC (National
Science and Engineering Research
Council of Canada) for his Ph.D. and one from ``Programme de Soutien de
Cotutelle de Th\`ese de doctorat du
Gouvernement du Qu\'ebec'' for his stay in Paris.
\bigskip
{\scap References}
\smallskip
\item{[1]} E.L. Ince, {\sl Ordinary differential equations}, Dover, New
York, 1956.
\item{[2]} B. Gambier, Acta Math. 33 (1910) 1.
\item{[3]} B. Ramani, A. Grammaticos and S. Lafortune, {\sl Schlesinger
Transformations for Linearisable
Equations}, Lett. Math. Phys., to appear.
\item{[4]} B. Grammaticos and A. Ramani, Physica A 223 (1995) 125.
\item{[5]} B. Grammaticos, A. Ramani and S. Lafortune, Physica A 253 (1998)
260.
\item{[6]} M.J. Ablowitz, A. Ramani, H. Segur, Lettere al Nuovo Cimento, 23
(1978) 333.
\item{[7]} A.S. Fokas and M.J. Ablowitz, J. Math. Phys. 23 (1982) 2033.
\item{[8]} C.M. Cosgrove and G. Scoufis, Stud. Appl. Math. 88 (1993) 25.
\end